# A theoretical approach for Pareto-Zipf law

Çağlar Tuncay
Department of Physics, Middle East Technical University
06531 Ankara, Turkey
caglart@metu.edu.tr

**Abstract:** We suggest an analytical approach for Pareto-Zipf law, where we assume random multiplicative noise and fragmentation processes for the growth of the number of citizens of each city and the number of the cities, respectively.

## 1 Introduction:

We show that the random multiplicative noise and the fragmentation processes [1] are conditionally similar and they give exponential variation within the related parameters with time. Hence, the probability of finding a city with a given population (size) decreases with this size and we have Pareto-Zipf law. The original theory is presented in the following section and the next one is devoted for discussion and conclusion.

## 2 Theory:

We define a uniformly distributed random number ($0 \leq \xi < 1$) which is utilized for several aims in the present simulation: For example, we use ($\xi_{I,t}$) at ($t$) for the city ($I$) with ($0 \leq \xi_{I,t} < 1$), etc. Population of the cities ($P_I(t)$) grow in time ($t$) with a random rate $R_I = R\xi_{I,t}$, where $R$ is universal within a *random multiplicative noise process*,

$$P_I(t) = (1 + R_I)P_I(t - 1) \quad . \tag{1}$$

Please note that, as the initial cities grow in population (size) they *fragment* in the meantime. [2] Eq (1) may be written as

$$P_I(t) = \prod_{t=1}^{t} (1 + R_I)P_I(1) \quad , \tag{2}$$

where $P_I(1)$ is the initial population of the city ($I$):

$$P_I(1) = \xi_{I,0} P_{max} \quad , \tag{3}$$

where ($P_{max}$) is the maximum of initial population that the ancestor cities may have and ($\xi_{I,0}$) is as in the first paragraph here. Thus, Eq (2) may be written as

$$P_I(t) = \prod_{t=1}^{t} (1 + R\xi_{I,t})\xi_{I,0} P_{max} \quad , \tag{4}$$

where ($R$) is the maximum for the growth rate of the cities (and similarly Eq. (1) becomes $P_I(t) = (1 + R\xi_{I,t})P_I(t - 1)$).

The number of the cities ($M(t)$) living at ($t$) increases with time in terms of fragmentation or emergence; and, the random fragmentation process can be conditionally related to the random multiplicative noise process.

## 2.1. Similarity between the random multiplicative noise and fragmentation processes

We assume ($M(t)$) to vary as

$$M(t) = (1 + E\xi_t)M(t - 1) \quad , \tag{5}$$

where ($E$) is the (net, i.e., after subtracting the maximum rate for the random extinction) maximum for the random emergence rate of the cities. Eq. (5) may be written as

$$M(t) = \prod_{t=1}^{t} (1 + E\xi_{,t})\xi_0 M_0 \quad , \tag{6}$$

where ($M_0$) is the number of ancestor cities. Please note that, the random fragmentation is the same as random extinction of the given city which is replaced by two cities which emerge newly at $t$ and the number of the current cities ($M(t)$) increases by 1, which is not important due to the present randomness. Suppose that the splitting ratio for the given fragmentation is $S$; this means that the offspring cities will have the following populations: $P_I(t)=SP_I(t)$ and $P_{M(t)+1}(t)=(1-S)P_I(t))$ both of which are clearly less than the population of the fragmented city. It is clear that the offspring cities may be interchanged; i.e., ($S$)→(1-$S$) with ($I$)→ ($M(t)+1$) and (1-$S$)→($S$) with ($M(t)+1$)→($I$). Thus a random fragmentation process may be considered as a random multiplicative noise process, where we have also a minimum for the population decay rate, since:

$$P_I(t)=SP_I(t)=S(1 + R\xi_{I,t})P_I(t - 1)) \tag{7}$$

which can be written as

$$P_I(t)=(S + SR\xi_{I,t})P_I(t - 1))=[1+(-1+S + SR\xi_{I,t})]P_I(t - 1) \tag{8}$$

or as

$$P_I(t)=[1-(1-S-SR\xi_{I,t})]P_I(t - 1) \quad , \tag{9}$$

where the factor ($1-S-SR\xi_{I,t}$) is the current (at ($t$)) random population decay rate and the maximum of which is ($1-S$) with $\xi_{I,t}=0$; and, similarly the mentioned maximum is ($S$) for ($M(t)+1$) with ($S$)→(1-$S$) or (1-$S$)→($S$) for ($I$) as mentioned within the text before Eq. (7).

Please note that, ($S=1/2$) may be taken as universal as well as a uniform random number, where average of ($S$) gives ½ in the long run. Secondly if ($S\approx 0$) or ($S\approx 1$) then it means that we do not have fragmentation practically, where one of the new cities goes extinction immediately.

## 2.2. Exponential growth in the population of each city and the number of the cities

Eq. (1) reads

$$P_I(t) - P_I(t - 1)= R_I P_I(t - 1) \quad , \tag{10}$$

which may be written as

$$\Delta P_I(t)/\Delta t = R_I P_I(t) \tag{11}$$

or

$$\Delta lnP_I(t) = R_I\Delta t \qquad (12)$$

and similarly for ($M(t)$) in Eq. (6), where $ln$ is the natural logarithm. Hence, the average of the logarithm of ($P_I(t)$) in Eq. (12) increases with $Rt/2$ in time ($t$) since ($R_I=R\xi_{I,t}$) and the average of the uniform numbers ($\xi_{I,t}$) between 0 and 1 is ½. Therefore, whatever the populations of the cities ($P_I(t)$) at a time $t$ are, the probability of finding a city with a population $P$ decreases exponentially with $P$ which gives the power law with exponent -1 (Pareto-Zipf law) as we show in the next section. A different derivation of more general power laws was given by Levy and Solomon [3].

*2.3. Finite sums over numerous exponential functions with random exponent (and independent random amplitude)*

Let us define uniformly distributed random numbers ($A_i$ and $B_j$) with $A_{min} \leq A_i < A_{max}$ and $B_{min} \leq B_j < B_{max}$, (i.e., $A_i=(A_{max}-A_{min})\xi_i+A_{min}$ and similarly for $B_j$) where each set is independent of each other. (We select $A_i$ and $B_j$ from different sets of random numbers, between which there is no connection.) Now we consider the exponential functions ($y_{ij}(t)$) for a real positive number $b$, where ($t$) is an integer variable with $t$=1, 2, 3, …, which may represent time or size, etc.

$$y_{ij}(t) = A_i \exp(-B_j b t) \quad . \qquad (13)$$

In Eq. (13) $A_i$ and $B_j$ may be fixed to be valid for all $t$'s utilized in the computation or they may be defined differently for each of the mentioned $t$'s. Please note that $y_{ij}(t)$ may describe the current number of the citizens living in a city ($P_I(t)$) or the current number of the cities ($M(t)$) at ($t$) after some suitable changes in the coefficient and the exponent. For example we may take $A_i$=1 and $B_j$=-$R_I$ with $b$=1 and we have $y_{ij}(t)=P_I(t)$ and similarly for $M(t)$. Here, we will utilize $y_{ij}(t)$ in Eq. (13) for describing the probability of finding a city with a population ($P$) which decreases exponentially with ($P$) as mentioned at the end of the previous section.

It is known that many physical functions show exponential decreasing distributions and exponential decays in time. Figure 1 is for one of many $y_{ij}(t)$ (Eq. (13)) with $b$=0.1 and $A_{min}=B_{min}$=0.0, $A_{max}=B_{max}$=1.0 (and hence, $A_i=\xi_i$ and $B_j=\xi_j$) which may be considered as the size distribution for the cities, where ($t$) describes population here. In Fig. 1 a new $B_j$ is tried for each of the utilized $t$'s along the horizontal axis which increase by 1 and the inset is the same as the figure, where $y_{ij}(t)$ is in the rank order and the vertical axis is logarithmic.

Now, we define

$$Y(t)=\sum_i^I \sum_j^J y_{ij}(t)/IJ \quad , \qquad (14)$$

with some big yet finite $I$ and $J$, where the order of sum is not important due to independence of the random numbers.

For homogeneously (yet, independent) distributions of the random numbers $A_i$ (for the amplitudes) and $B_i$ (for the exponents) one may utilize the general theorem of central limit with large $N$ or just convert the double sum into a double integral over $A$ ($A_i \rightarrow A$) and $B$ ($B_j \rightarrow B$), which vary linearly (since the random numbers are homogeneously distributed) between the extrema to obtain the following equalities:

$$Y(t)=\sum_i^I \sum_j^J y_{ij}(t)/IJ$$
$$\propto -(A^2_{max}-A^2_{min})[\exp(-B_{max}bt)-\exp(-B_{min}bt)]/(2bt) \quad , \qquad (15)$$

and, for $B_{min}$=0, we have (for 1«$2bt$)

$$Y(t) \propto [(A^2_{max} - A^2_{min})/2b] \, t^{-1} \, , \qquad (16)$$

since, the exponential term (Eq. (15)) approaches to zero as $t$ increases. And the result is power law with exponent minus one. Figure 2 displays $Y(t)$ (Eq. (14)) with $I=J=1000$, for $A_{min}= B_{min}=0.0$, $A_{max}=B_{max}=1.0$, where we have no fluctuation in $Y$ for the random numbers selected as the same for each of the utilized $t$'s within the simulation (thick line). And fluctuation in $Y$ increases in magnitude, as $t$ increases for the random numbers selected for each of the utilized $t$'s within the simulation differently. And, in both cases we have power law; as the arrow with slope -1 indicates in log-log scales. Obviously, the important issue in Fig. 2 is that $Y(t)$ is a power function for large ($t$) with the power of (about) minus unity, i.e.,

$$Y(t) = \sum_i^I \sum_j^J y_{ij}(t)/IJ \propto t^{-1} \, , \qquad (17)$$

which is similar to Pareto-Zipf law (thin line), if ($t$) stands for the city population.

**3 Discussion and Conclusion:**

$B_{max}$ may always be taken as unity, and $b$ may be varied accordingly in Eqs. (13)-(17). For $A_{min}=0$ and $A_{max}=1$ (since $A_i=1$ may be taken), $Y(t)$ in Eq. (16) may give the distribution of the number of cities over the number of the citizens with $t \rightarrow P$ and $b=R$. It is clear that, $R$ and $t$ may not be independent if the individual populations ($P_I(t)$) should repeat (or approximate) the real data. Secondly, the random fragmentation or the multiplicative noise process with a negative growth (i.e., decay) rate run the cities to extinction, where the individual population ($P_I(t)$) falls to zero. We may state that, our analysis is general for any set of independent random numbers for amplitudes (provided all of the limits $A_{max}$ and $A_{min}$, etc., are finite) and that for exponents (provided $B_{min}= 0$, where $B_{max}=1$ may be taken after changing $b$ accordingly). The present approach may be followed also for the biological speciation and extinction, where the lifetimes of the families within a specified order maybe exponentially distributed (due to the randomness of extinctions as well as the distinctive different life times between the different order, which is known as Van Valen's Red Queen hypothesis). And then, the species should be distributed following a power law with power (about) minus unity as Eqs. (13)-(17) imply. [2]

The important points are: 1) The initial parameters ($P_{max}$ and $M_0$) in Eqs. (4) and (7), respectively are not important for the results and changing them amounts to shifting (back or forth) the origin for the time axis (for example in Eqs. (4) and (7)). 2) Changing the rates ($R$, $E$ and $S$ in Eqs. (4), (7) and (8), respectively amounts to scaling the time axis, i.e., changing the unit for time accordingly. And, in terms of 1) and 2) here, it is clear that the mentioned theoretical data displays "universality" since they may be collapsed on to a single line. 3) The same reason, i.e., the probability of finding a city for a given population is decreasing exponentially with size gives exponential decay in the life times, since the populations grow exponentially in time. This prediction may be considered as a "natural election" for the cities: The cities either go extinct quickly or live long.

**ACKNOWLEDGEMENT**

The author is thankful to Dietrich Stauffer who will retire soon, for his critical reading and corrections, valued discussions, inspirations, suggesting references, and many other contributions to the papers of the author.

**FIGURES**

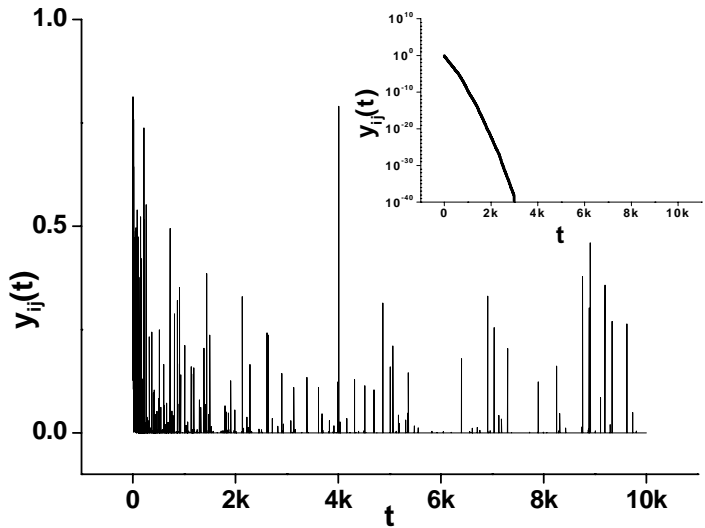

**Figure 1**     $y_{ij}(t)$ (Eq. (14)) with b=0.1 and $A_{min} = B_{min} = 0.0$, $A_{max} = B_{max} < 1.0$, which may be considered as the population of cities, etc., where a new $B_j$ is tried for each of the utilized ($t$)'s along the horizontal axis which increase by 1. The inset is the same as the figure, where $y_{ij}(t)$ is in the rank order and the vertical axis is logarithmic.

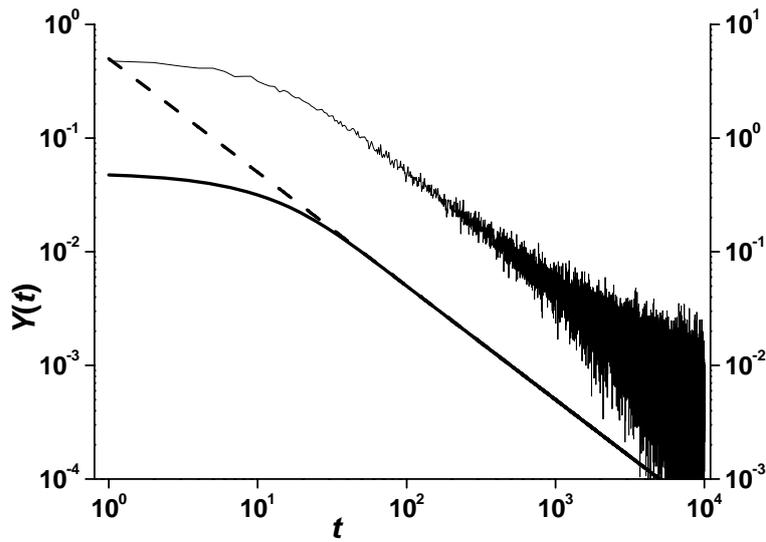

**Figure 2**     $Y(t)$ with $I=J=1000$, for $A_{min}=B_{min}=0.0$, $A_{max}=1.0$, $B_{max}=1.0$, where the oscillating plot is for Eq. (14) with the vertical axis on left. The solid plot is for the analytical expression in Eq. (15) and the dashed line is for Eq. (16) both with the vertical axis on right. The shifted vertical axes have the same units.